\documentclass[reprint,showpacs,amsmath,amssymb,aps,pre]{revtex4-1}
\usepackage{dcolumn}
\usepackage{bm}
\usepackage[dvipdfmx]{graphicx}
\usepackage[dvipdfmx]{hyperref}
\usepackage[mathlines]{lineno}

\begin{document}

\title{Thermal transport in disordered harmonic chains revisited: Formulation of thermal conductivity and local temperatures}
\author{Kiminori Hattori and Shohei Kumatoriya}
\affiliation{Department of Systems Innovation, Graduate School of Engineering Science, Osaka University, Toyonaka, Osaka 560-8531, Japan}
\date{\today}

\begin{abstract}	
In this paper, thermal transport in bond-disordered harmonic chains is revisited in detail using a nonequilibrium Green's function formalism.
For strong bond disorder, thermal conductivity is independent of the system size.
However, kinetic temperatures described by the local number of states coupling to external heat reservoirs are anomalous since they form a nonlinear profile in the interior of the system.
Both results are accounted for in a unified manner in terms of the frequency-dependent localization length.
From this argument, we derive a generic formula describing the asymptotic profile of local temperatures in a disordered harmonic chain.
A linear temperature profile obeying Fourier's law is recovered by contact with a self-consistent reservoir of Ohmic type even in the limit of weak system-reservoir coupling.
This verifies that mechanisms leading to local thermal equilibrium and breaking total momentum conservation are essential for Fourier transport in low dimensions.
\end{abstract}

\maketitle

\section{Introduction}
\label{sec:1}

Fourier's law is a celebrated phenomenological law that relates heat current to a temperature gradient as ${\mathbf{J}} = -\kappa\nabla\theta$, where $\kappa$ is the material-dependent thermal conductivity.
Because of energy conservation, this law predicts a linear temperature profile for a small temperature bias $\Delta\theta$ along the direction of heat flow in the steady state.
It also follows that under a fixed bias, the heat current varies as $J \propto N^{-1}$, where $N$ denotes the system size.
In nonequilibrium statistical physics, it is a fundamental challenge to derive Fourier's law from first principles.
This issue remains unresolved despite extensive theoretical studies thus far.
From these studies, it is widely accepted at present that Fourier's law is genuinely broken in a low-dimensional lattice system without external forces that break total momentum conservation \cite{ref:1,ref:2,ref:3, ref:4, ref:5}.
In particular, anharmonic Fermi-Pasta-Ulam chains and disordered harmonic chains are the typical examples showing this anomaly.
For these systems, the finite-size thermal conductivity defined as $\kappa _N = JN/\Delta\theta$ diverges in the thermodynamic limit $N \to \infty $.
The predicted anomalous non-Fourier transport is observed experimentally in carbon and boron-nitride nanotubes \cite{ref:6} and in single-layer graphene \cite{ref:7}.
In experiments, dimensional crossover of thermal transport is also seen in few-layer graphene \cite{ref:8}.

Harmonic chains with quenched disorders constitute an integrable model that enables analytical and numerical treatments in an exact manner.
As mentioned above, the relevant models have been intensively explored over many decades to elucidate heat transport occurring in low dimensions \cite{ref:2, ref:3, ref:4, ref:9, ref:10, ref:11, ref:12, ref:13, ref:14, ref:15}.
The previous studies, which generally assume spatially uncorrelated mass disorder, support anomalous transport in momentum-conserving systems.
Nonetheless, $\kappa _N$ scaling normally with $N$ has been found in recent years for particular classes of disorders such as correlated mass disorder \cite{ref:16} and uncorrelated bond disorder \cite{ref:17, ref:18}.
The recovery of normal conductivity despite total momentum conservation seems to disprove the prevailing conjecture, if the normality is also verified for local temperatures in the interior of the system.

The temperature profile in a disordered harmonic chain has been little studied \cite{ref:2, ref:4, ref:19}.
The previous studies in this direction are limited to dealing with a relatively small system of $N \sim 100$.
In the presence of quenched disorder, most of the nonzero frequency modes are exponentially localized and thereby it takes an extremely long time to reach the nonequilibrium steady state.
This feature tends to disable conventional stochastic approaches \cite{ref:2, ref:4}.
Thus, the asymptotic profile of local temperatures remains an open problem for disordered chains.

In this paper, we address the following two issues for thermal transport in disordered harmonic chains.

First, we examine whether Fourier transport is genuinely reproducible in disordered chains.
To this end, we employ the nonequilibrium Green's function (NEGF) formalism \cite{ref:20, ref:21, ref:22} or equivalently the quantum Langevin equation (QLE) formalism \cite{ref:3, ref:4, ref:23}.
Following these formalisms, stationary heat current flowing in a two-terminal system subjected to a temperature bias is described by the Landauer-B\"{u}ttiker (LB) formula.
In this nonequilibrium situation, kinetic temperatures are formulated in terms of the local number of states coupling to external heat reservoirs attached to the system.
 We show that the local temperatures are anomalous even in a system exhibiting normal conductivity.
 This result is quantitatively accounted for by a generic formulation of the asymptotic profile of local temperatures in a disordered harmonic chain.
 
In the two-terminal geometry, a long enough disordered chain is not expected to locally equilibrate since no energy transfer is allowed between localized and extended states.
 This motivates us to consider what role is played by local equilibration in heat transport.
 We address this second issue by attaching a self-consistent reservoir (SCR) to the system to enforce local equilibration \cite{ref:23, ref:24, ref:25, ref:26, ref:27, ref:28, ref:29, ref:30}.
 We show that a normal temperature profile is recoverable by contact with an Ohmic SCR even in the limit of weak system-reservoir coupling, verifying that mechanisms leading to local thermal equilibrium and breaking total momentum conservation are necessary for Fourier transport in low dimensions.
 
The paper is organized as follows.
In Sec. \ref{sec:2}, we formulate thermal current and kinetic temperatures in the two-terminal system.
In Sec. \ref{sec:3}, we elucidate the two issues given above by using numerical calculations.
Finally, Sec. \ref{sec:4} provides a summary.

\section{Theoretical formulation}
\label{sec:2}

Throughout this paper, we shall work in units where $\hbar = k_B = 1$.
We consider a one-dimensional lattice with nearest-neighbor harmonic interactions.
The lattice Hamiltonian is written as
\begin{equation*}
H = \frac{1}{2}(\sum\limits_j {{m_j}{{\dot q}_j}{\dot q}_j} + \sum\limits_{jj'} {{K_{jj'}}{q_j}{q_{j'}}}) ,
\end{equation*}
where ${q_j}$ denotes the displacement of a particle of mass ${m_j}$ from its equilibrium position.
This Hamiltonian allows for random masses $\{ {m_j}\} $ giving rise to diagonal disorder.
In general, the force constant matrix has the symmetry $K_{jj'} = K_{j'j}$, and obeys the acoustic sum rule ${\sum _j}{K_{jj'}} = 0$ because of translational invariance \cite{ref:31}.
It is easy to see that this constraint ensures the conservation of total momentum as well as the persistence of the fundamental zero mode.
For instance, the sum rule is maintained for the coupling matrix of the form
\begin{equation*}
{K_{jj'}} = ({k_j} + {k_{j - 1}}){\delta _{jj'}} - {k_j}{\delta _{j',j + 1}} - {k_{j - 1}}{\delta _{j',j - 1}} ,
\end{equation*}
which describes off-diagonal disorder where one regards bond strengths $\{ {k_j}\} $ as random variables.

To analyze thermal transport, we assume a standard two-terminal setup, where the central system consisting of $N$ lattice sites is coupled at both ends to two leads denoted as $L$ and $R$, which serve as heat reservoirs sustained at unequal temperatures ${\theta _L}$ and ${\theta _R}$, respectively.
In this setup, the energy flux is described by the LB formula, which can be derived from the NEGF formalism \cite{ref:20, ref:21, ref:22} as well as the QLE formalism \cite{ref:3, ref:4, ref:23}.
In terms of this formula, heat current flowing in the two-terminal system is expressed as $J = G({\theta _L} - {\theta _R})$ for a small enough temperature bias ${\theta _L} - {\theta _R}$.
The thermal conductance is defined as
\begin{equation}
\label{eq:1}
G = \frac{1}{{2\pi }}\int_0^\infty {d\omega \frac{{\partial f}}{{\partial \theta }}\omega T(\omega )} ,
\end{equation}
where $T = {\Gamma _L}{\Gamma _R}{\left| {{g_{1N}}} \right|^2}$ is the transmission coefficient, $g_{jj'}$ is the retarded Green's function between sites $j$ and $j'$, $\Gamma _\nu$ is the linewidth function for lead $\nu \in \{ L,R\} $, and $f = {({e^{\omega /\theta }} - 1)^{-1}}$ is the Bose function for phonons.
The finite-size thermal conductivity is given by $\kappa _N = G(N+1)$ for the system of size $N$.

The velocity-velocity correlation function $\left\langle {{{\dot q}_j}{{\dot q}_j}} \right\rangle$ is also derivable from the NEGF formalism or equivalently the QLE formalism \cite{ref:23}.
The equal-time correlation leads to the local kinetic temperature formulated as
\begin{equation}
\label{eq:2}
{\tilde \theta _j} = {m_j}\left\langle {{{\dot q}_j}{{\dot q}_j}} \right\rangle = \sum\limits_{\nu = L,R} {\int_0^\infty {d\omega ({f_\nu } + \frac{1}{2})\omega {D_{j\nu }}} } ,
\end{equation}
where ${D_{j\nu }} = {m_j}\omega {\Gamma _\nu }{\left| {{g_{j\nu }}} \right|^2}/\pi$ and ${f_\nu } = f({\theta _\nu })$.
In this notation, the index $\nu$ for $g_{j\nu }$ is assigned to the internal site connected to terminal $\nu $, i.e., ${g_{jL}} \equiv {g_{j1}}$ and ${g_{jR}} \equiv {g_{jN}}$.
See, Appendix \ref{appendix:A} for the derivation of Eq. (\ref{eq:2}).
Linearizing Eq. (\ref{eq:2}), one obtains
\begin{equation}
\label{eq:3}
{\tilde \Theta _j} = \frac{{{{\tilde \theta }_j} - \tilde \theta _j^{(0)}}}{{{\theta _L} - {\theta _R}}} = \sum\limits_{\nu = L,R} {{\Theta _\nu }\int_0^\infty {d\omega \frac{{\partial f}}{{\partial \theta }}\omega {D_{j\nu }}} } .
\end{equation}
Here, ${\Theta _\nu } = ({\theta _\nu } - \theta )/({\theta _L} - {\theta _R})$ and $\theta = ({\theta _L} + {\theta _R})/2$.
The dimensionless terminal temperature satisfies ${\Theta _L} = - {\Theta _R} = 1/2$.
The quantity $\tilde \theta _j^{(0)}$ denotes the kinetic temperature at zero bias ${\theta _L} = {\theta _R} = \theta $.

In the high-temperature limit $\theta \to \infty $, the factor $\tfrac{{\partial f}}{{\partial \theta }}\omega $ approaches unity irrespective of $\omega $ so that Eq. (\ref{eq:1}) is reduced to
\begin{equation}
\label{eq:4}
G = \frac{1}{{2\pi }}\int_0^\infty {d\omega T(\omega )} .
\end{equation}	
Similarly, Eq. (\ref{eq:3}) becomes
\begin{equation}
\label{eq:5}
{\tilde \Theta _j} = \sum\limits_{\nu = L,R} {{\mathcal{N}_{j\nu }}{\Theta _\nu }} ,
\end{equation}	
in this limit, where ${\mathcal{N}_{j\nu }} = \smallint _0^\infty d\omega {D_{j\nu }}$.
For an isolated system, the local density of states per site is given by ${D_j} = - 2{m_j}\omega \operatorname{Im}{g_{jj}}/\pi $, and hence ${\mathcal{N}_j} = \smallint _0^\infty d\omega {D_j}$ denotes the local number of states.
Following the identity $ - 2\operatorname{Im}{g_{jj}} = {\sum _\nu }{\Gamma _\nu }{\left| {{g_{j\nu }}} \right|^2}$, we observe ${\mathcal{N}_j} = {\sum _\nu }{\mathcal{N}_{j\nu }}$ for the system in contact with leads.
Thus, ${\mathcal{N}_{j\nu }}$ represents the contribution of lead $\nu $ to ${\mathcal{N}_j}$.
In what follows, we refer to ${\mathcal{N}_{j\nu }}$ as the local number of states coupling to lead $\nu $, since ${g_{j\nu }}$ incorporated in ${\mathcal{N}_{j\nu }}$ describes the correlation between $j$ and $\nu $.

At zero bias, the kinetic temperature is calculated from Eq. (\ref{eq:2}) to be
\begin{equation}
\label{eq:6}
\tilde \theta _j^{(0)} = {\mathcal{N}_j}\theta ,
\end{equation}
in the limit of $\theta \to \infty $.
This yields a compact general expression
\begin{equation}
\label{eq:7}
{\tilde \theta _j} = \sum\limits_{\nu = L,R} {{\mathcal{N}_{j\nu }}{\theta _\nu }} ,
\end{equation}
which is valid for an arbitrary small bias including zero bias.
Equations (\ref{eq:5})-(\ref{eq:7}) are a set of simple and novel formulas relating the kinetic temperatures and the local number of states.
The zero-bias result, Eq. (\ref{eq:6}), disagrees with a na\"{i}ve expectation that $\tilde \theta _j^{(0)} = \theta $ unless ${\mathcal{N}_j} = 1$.
It is shown in the literature that the latter is valid only in the limit of weak system-reservoir coupling \cite{ref:2, ref:4, ref:9, ref:32} and does not necessarily hold for generic open systems acting as heat conductors \cite{ref:32, ref:33}.
See, Appendix \ref{appendix:B} for $\mathcal{N}_j$ in the weak-coupling limit.

\section{Numerical results and discussion}
\label{sec:3}

In this study, random variables $X \in \{ {m_j}/m,{k_j}/k\} $ are assumed to follow uniform distributions over the intervals $X \in [0.5,1.5]$ for weak disorder and $X \in (0,2]$ for strong disorder, where $m$ and $k$ denote the expected values of ${m_j}$ and ${k_j}$, respectively.
Thus, we address four types of uncorrelated disorders consisting of weak mass disorder (WMD), weak bond disorder (WBD), strong mass disorder (SMD), and strong bond disorder (SBD).
In numerical calculations, disorder averaging is performed over ${10^4}$ random configurations unless stated otherwise.
Note that the uniform distribution describing strong disorder is equivalent to the power-law distribution $P(X) \propto {X^{\epsilon - 1}}$ with $\epsilon = 1$, which was employed in the previous study \cite{ref:17, ref:18}.

The retarded Green's function of a linear chain consisting of $n$ lattice sites is analyzable in the recursive manner \cite{ref:4, ref:21, ref:29, ref:34, ref:35} formulated as
\begin{eqnarray*}
{g_{nn}} & = & {(s_n^{ - 1} - {K_{n,n - 1}}{g_{n - 1,n - 1}}{K_{n - 1,n}})^{ - 1}} ,\\
{g_{1n}} & = & {g_{1,n - 1}}{K_{n - 1,n}}{g_{nn}} ,	
\end{eqnarray*}
where ${s_n} = ({m_n}{\omega ^2} - {K_{nn}}){}^{ - 1}$ represents the Green's function of an isolated site.
In the two-terminal geometry, the Green's function ${g_{1N}}$ joining two ends is computed by adding sites one by one to an isolated lead until reaching the opposite lead.
The lead is assumed to be a semi-infinite ordered chain with equal particles of mass $m$ and equal bond strength $k$, which coincide with the expected values of $m_j$ and $k_j$ in the system, respectively.
The coupling to leads induces the self-energy ${\Sigma _L} = {\Sigma _R} = \bar \Sigma $ in the system, which is analytically derived to be $\bar \Sigma = - k{e^{2i\sin^{-1}z}}$, where $z = \omega /2t$ and $t = \sqrt {k/m} $.
Note that $2t$ corresponds to the bandwidth of the ordered chain.
In the recursive Green's function formalism, the self-energy is incorporated into ${g_{11}}$ and ${g_{NN}}$ at the boundary sites.
The associated linewidth ${\Gamma _L} = {\Gamma _R} = \bar \Gamma $ is given by $\bar \Gamma = - 2\operatorname{Im}\bar \Sigma $.
Following an analogous procedure, the internal two-point correlations ${g_{j1}}$ and ${g_{jN}}$ are derived numerically.
See, Ref. \cite{ref:35} for more details.
The present model constitutes an infinitely extended chain.
The calculation assumes the free boundary condition to fulfill the acoustic sum rule in the entire system.
Otherwise, total momentum conservation is violated.

In the following, we elucidate the two issues raised in Sec. \ref{sec:1} by means of numerical calculations.
All the numerical results shown below are obtained in the high-temperature limit $\theta \to \infty $.
The features expected at finite temperatures are discussed at the end of Sec. \ref{sec:3}.
First, we address the validity of Fourier transport in disordered chains.

\begin{figure}
\centering
\includegraphics{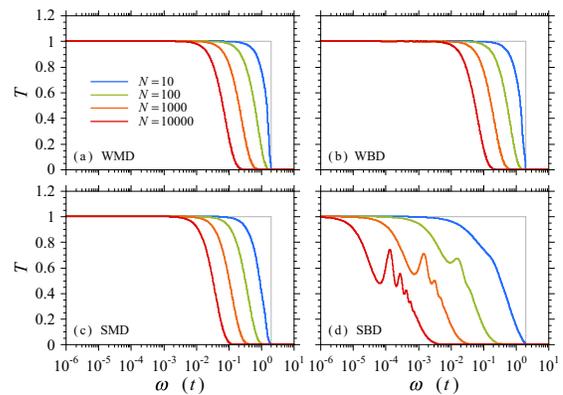}
\caption{Transmission coefficient $T$ as a function of frequency $\omega $. Four panels display the numerical results for (a) WMD, (b) WBD, (c) SMD and (d) SBD. In each panel, the system size is varied as $N = 10$, 100, 1000 and 10000. The thin gray line shows $T$ in the absence of disorder as a reference.}
\label{fig:1}
\end{figure}

Figure \ref{fig:1} summarizes the transmission $T(\omega )$ for four types of disorders.
In general, $T(\omega )$ is invariable and fixed at unity in the $\omega \to 0$ limit.
This indicates that the relevant low-frequency modes are delocalized and immune to disorder by virtue of the acoustic sum rule.
At high frequencies, $T(\omega )$ tends to vanish, showing that the relevant high-frequency modes are localized and no longer carry heat current across the system.
The cutoff frequency separating these two regimes lowers with increasing the size of the chain $N$.
For SBD, interference fringes are formed in the transitional regime.
This is due to a boundary mismatch in the present model, where disorder is absent in the leads attached to the system.
The occurrence of interference also manifests coherent heat transport.

\begin{figure}
\centering
\includegraphics{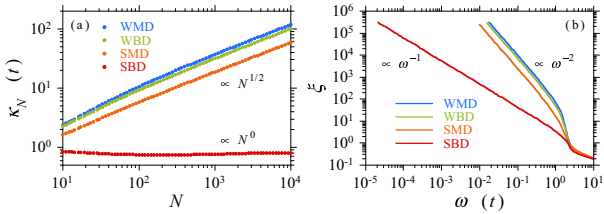}
\caption{(a) Finite-size thermal conductivity ${\kappa _N}$ as a function of system size $N$ and (b) localization length $\xi $ as a function of frequency $\omega $ for WMD, WBD, SMD and SBD.}
\label{fig:2}
\end{figure}

The finite-size conductivity ${\kappa _N}$ is displayed in Fig. \ref{fig:2}(a).
As shown in the figure, ${\kappa _N}$ behaves as ${\kappa _N} \propto {N^0}$ for SBD and ${\kappa _N} \propto {N^{1/2}}$ for the other types of disorders.
These power-law behaviors correlate to the localization length $\xi (\omega )$ as explained below.
For exponential localization, the correlation function follows
\begin{equation}
\label{eq:8}
\lim_{\left| {j - j'} \right| \to \infty } \left| {{g_{jj'}}(\omega )} \right| \propto {e^{ - \left| {j - j'} \right|/\xi (\omega )}} ,
\end{equation}
and hence $T(\omega ) \approx {e^{ - 2(N - 1)/\xi (\omega )}}$ in the asymptotic limit $N \to \infty $.
Accordingly, $\xi (\omega )$ is numerically derived from $T(\omega )$ via $\xi (\omega ) = - \lim_{N \to \infty } \tfrac{{2(N - 1)}}{{\ln T(\omega )}}$.
The numerical results obtained for $N \approx {10^8}$ after averaging over 10 disorder realizations are shown in Fig. \ref{fig:2}(b).
As seen in the figure, $\xi (\omega )$ varies as ${\omega ^{ - \alpha }}$ at low enough frequencies.
The power-law exponent is found to be $\alpha = 1$ for SBD and $\alpha = 2$ for the other types of disorders.
It may be worth noting that in the second-order Born approximation, particle mass and bond strength are renormalized by disorders such that $m \to m(1 + i\tfrac{{\sigma _m^2}}{{{m^2}}}\tfrac{\omega }{{2t}})$ and $k \to k(1 - i\tfrac{{\sigma _k^2}}{{{k^2}}}\tfrac{\omega }{{2t}})$, where $\sigma _x^2$ refers to the variance of a random variable $x$.
These renormalized parameters lead to ${\xi ^{ - 1}}(\omega ) = (\tfrac{{\sigma _m^2}}{{{m^2}}} + \tfrac{{\sigma _k^2}}{{{k^2}}}){(\tfrac{\omega }{{2t}})^2}$ in the limit of $\omega \to 0$.
This relation agrees with the estimate from a renormalization group study \cite{ref:36, ref:37}, and accounts for the numerical results shown in Fig. \ref{fig:2}(b) except for SBD.
For SBD, $\alpha = 1$ is shown from a scaling argument in the literature \cite{ref:38}.
This prediction coincides with the observation.
In terms of the localization length scaling as $\xi (\omega ) \propto {\omega ^{ - \alpha }}$, the high-frequency cutoff amounts to ${\omega _N} \propto {N^{ - 1/\alpha }}$.
In the $\theta \to \infty $ limit, $G = {\omega _N}/2\pi $ so that we arrive at
\begin{equation}
\label{eq:9}
{\kappa _N} \propto {N^{(\alpha - 1)/\alpha }} ,
\end{equation}
in the asymptotic regime \cite{ref:3, ref:4}.
This explains the numerical results shown in Fig. \ref{fig:2}(a).

\begin{figure}
\centering
\includegraphics{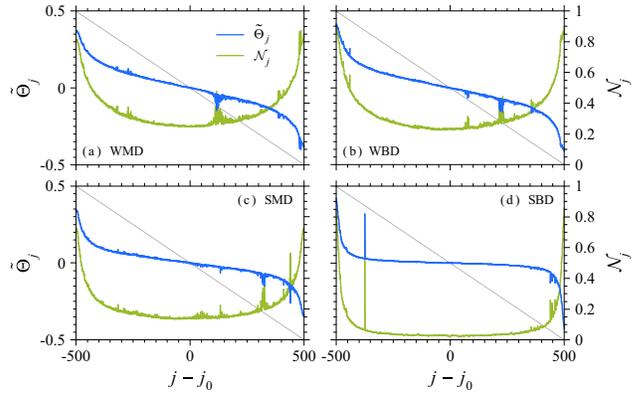}
\caption{Kinetic temperature ${\tilde \Theta _j}$ and local number of states ${\mathcal{N}_j}$ for $N = 1000$. Four panels display the numerical results for (a) WMD, (b) WBD, (c) SMD and (d) SBD. The reference point ${j_0} = (N + 1)/2$ is used to symmetrize these plots. The thin gray line shows the temperature profile obeying Fourier's law as a reference.}
\label{fig:3}
\end{figure}

\begin{figure}
\centering
\includegraphics{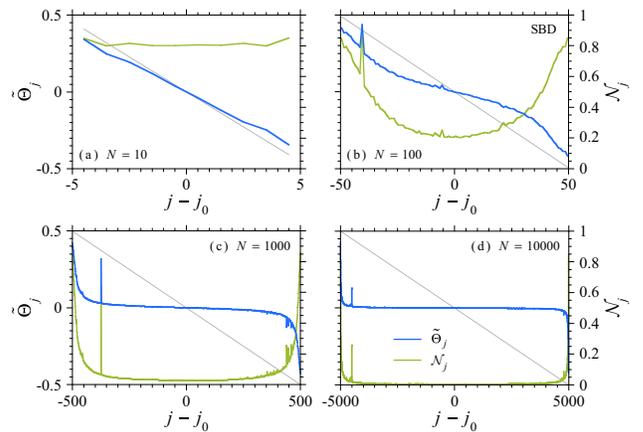}
\caption{Kinetic temperature ${\tilde \Theta _j}$ and local number of states ${\mathcal{N}_j}$ for SBD. Four panels display the numerical results for (a) $N = 10$, (b) 100, (c) 1000 and (d) 10000. The reference point ${j_0} = (N + 1)/2$ is used to symmetrize these plots. The thin gray line shows the temperature profile obeying Fourier's law as a reference.}
\label{fig:4}
\end{figure}

Figure \ref{fig:3} displays the local kinetic temperature ${\tilde \Theta _j}$ and the local number of states ${\mathcal{N}_j}$ in the two-terminal system of size $N = 1000$.
It is clearly seen in the figure that ${\tilde \Theta _j}$ is nonlinear and ${\mathcal{N}_j}$ is nonuniform for all types of disorders.
Recall that ${\mathcal{N}_j} = {\sum _\nu }{\mathcal{N}_{j\nu }}$ and ${\mathcal{N}_{j\nu }}$ relates to the correlation function ${g_{j\nu }}$.
Therefore, a salient reduction of ${\mathcal{N}_j}$ observed in the bulk implies that strongly localized states in the bulk are decoupled from the leads due to vanishingly weak correlations between them.
In view of this, it is naturally expected that both ${\tilde \Theta _j}$ and ${\mathcal{N}_j}$ tend to vanish as $N \to \infty $ in the bulk of a disordered chain as a result of Anderson localization.
The expected behavior is confirmed in Fig. \ref{fig:4}, where the numerical results are shown for various $N$ in the case of SBD.
The physical origin of a few sharp peaks seen in these numerical results is not entirely clear, although it is likely that they stem from resonant interference accidentally occurring inside the system in particular disorder configurations.
It has been found from repeated calculations that they have an essentially random nature.
In view of this, it is expected that these noise-like peaks disappear after averaging over infinitely many disorder realizations.

\begin{figure}
\centering
\includegraphics{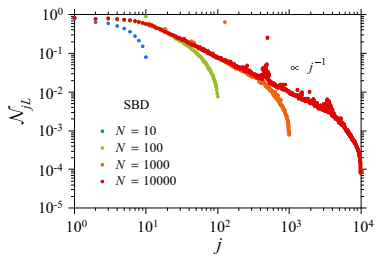}
\caption{Local number of states ${\mathcal{N}_{jL}}$ coupling to lead $L$ for SBD. The system size is varied as $N = 10$, 100, 1000 and 10000.}
\label{fig:5}
\end{figure}

The kinetic temperatures in the bulk of a disordered chain are more quantitatively analyzable.
The approach to this problem is parallel to that leading to Eq. (\ref{eq:9}).
It is shown from Eq. (\ref{eq:8}) that ${D_{jL}} \propto {e^{ - 2(j - 1)/\xi }}$ and ${D_{jR}} \propto {e^{ - 2(N - j)/\xi }}$. Hence, one finds that ${\mathcal{N}_{jL}} \propto {(j - 1)^{ - 1/\alpha }}$ and ${\mathcal{N}_{jR}} \propto {(N - j)^{ - 1/\alpha }}$ for $\xi (\omega ) \propto {\omega ^{ - \alpha }}$.
Then, Eq. (\ref{eq:7}) results in
\begin{equation}
\label{eq:10}
{\tilde \theta _j} \propto {(j - 1)^{ - 1/\alpha }}{\theta _L} + {(N - j)^{ - 1/\alpha }}{\theta _R} ,
\end{equation}
which describes the asymptotic profile of local temperatures.
It is easy to see that the normalized temperature obeys ${\tilde \Theta _j} \propto {(j - 1)^{ - 1/\alpha }} - {(N - j)^{ - 1/\alpha }}$ in terms of Eq. (\ref{eq:5}).
Equation (\ref{eq:10}) is a main result of the present study.
Note that ${{\tilde \theta }_j}$ and ${\mathcal{N}_{j\nu }}$ discussed here are relevant to the subsystem consisting of delocalized states which participate in heat transport.
It can be seen from Eq. (\ref{eq:10}) that the equality $\tilde \theta _j^{(0)} = \theta $ no longer holds in the present model.
Figure \ref{fig:5} shows ${\mathcal{N}_{jL}}$ computed for various $N$ in the case of SBD.
The local number of states coupling to lead $L$ exhibits a power-law decay as ${j^{ - 1}}$ in the range $1 \ll j < N/2$.
This corroborates the above argument since $\alpha = 1$ for SBD.
For the other types of disorders, numerical results do not contradict $\alpha = 2$ (not shown).
On the other hand, ${\mathcal{N}_{jL}}$ rapidly decreases in the vicinity of the opposite boundary at $j = N$.
It is presumed that this phenomenon can be ascribed to the mismatched boundary, although the associated mechanism is not fully clear at the present stage of investigation.

In stationary situations, heat current should be constant throughout a disordered chain owing to energy conservation.
In the present model, the expected uniformity has been verified by computing local bond currents ${J_{jj'}} = \operatorname{Re}{K_{jj'}}\left\langle {{{\dot q}_j}{q_{j'}}} \right\rangle $ \cite{ref:23, ref:30}.
Nevertheless, the profiles of kinetic temperatures are strongly nonlinear in the asymptotic regime for all types of disorders.
In view of this, Fourier's law is broken in the thermodynamic limit even in the system exhibiting normal conductivity.

In this paper, we restrict our attention to uncorrelated disorders.
It is shown in the literature \cite{ref:16} that a long-range spatial correlation of random masses leads to a localization length with a power-law exponent $\alpha = 1 + \delta $ and hence the finite-size conductivity scales as ${\kappa _N} \propto {N^{\delta /(1 + \delta )}}$.
The factor $\delta $ is positive, but arbitrarily small, so that normal scaling is recoverable for the correlated disorder in the limit of $\delta \to 0$.
However, the present theory predicts a nonlinear profile of kinetic temperatures even in this case.
Therefore, the conclusion stated above is not altered.

Next, we consider the SCR model to clarify the second issue.
In this model, the system is in contact with fictitious stochastic reservoirs, which bring about decoherence in the system following the fluctuation-dissipation theorem.
The self-energy due to the inner reservoir is describable by $\Sigma (\omega ) = - im\gamma \operatorname{sgn} (\omega ){\left| \omega \right|^\beta }$, where $\beta $ is an arbitrary positive real.
Note that this form satisfies the reality condition $\Sigma (\omega ) = {\Sigma ^*}( - \omega )$.
The SCR model usually arranges such fictitious reservoirs at all sites of the system \cite{ref:23, ref:24, ref:25, ref:26, ref:27, ref:28, ref:29, ref:30}.
In this study, we employ a simpler model consisting only of a single reservoir that contacts only a single site $j$.
Local equilibration between the system and the inner reservoir coupling to the site $j$ is attained under the self-consistent condition that no net energy current flows into the reservoir held at a certain temperature ${\theta _j}$.
This adiabatic condition is expressed as ${\sum _\nu }{G_{j\nu }}({\theta _j} - {\theta _\nu }) = 0$ in the multiterminal LB formalism.
Here, the internal conductance ${G_{j\nu }}$ is defined similarly as the two-terminal conductance $G$, i.e., ${G_{j\nu }} = \tfrac{1}{{2\pi }}\smallint _0^\infty d\omega {T_{j\nu }}(\omega )$ in the $\theta \to \infty $ limit, where ${T_{j\nu }} = \Gamma {\Gamma _\nu }{\left| {{g_{j\nu }}} \right|^2}$ and $\Gamma = - 2\operatorname{Im}\Sigma $.
In terms of ${G_{j\nu }}$, the local temperature ${\theta _j}$ probed in our SCR setup is simply formulated as
\begin{equation}
\label{eq:11}
{\theta _j} = \frac{{\sum\limits_{\nu = L,R} {{G_{j\nu }}{\theta _\nu }} }}{{\sum\limits_{\nu = L,R} {{G_{j\nu }}} }} .
\end{equation}
The probe temperature can be rewritten as
\begin{equation}
\label{eq:12}
{\Theta _j} = \frac{{{\theta _j} - \theta }}{{{\theta _L} - {\theta _R}}} = \frac{{\sum\limits_{\nu = L,R} {{G_{j\nu }}{\Theta _\nu }} }}{{\sum\limits_{\nu = L,R} {{G_{j\nu }}} }} ,
\end{equation}
in the dimensionless fashion.
Note that $\theta _j^{(0)} = \theta $ at zero bias ${\theta _L} = {\theta _R} = \theta $.
This property is distinct from that for the kinetic temperature $\tilde \theta _j^{(0)}$ derived in the absence of SCR.
Equations (\ref{eq:11}) and (\ref{eq:12}) are valid even in the weak-coupling limit $\gamma \to 0$.
In this limit, ${G_{j\nu }}$ is linear in $\gamma $ so that ${\theta _j}$ is independent of $\gamma $.
It is easy to see that the effective two-terminal conductance is expressed as $G + {G_{jL}}{G_{jR}}/({G_{jL}} + {G_{jR}})$ for the present SCR model.
The second term related to SCR is linear in $\gamma $ and vanishes in the weak-coupling limit, implying that scattering and dephasing due to SCR are negligible in this limit.
It should be emphasized that even in this limit, we consider an extremely small but nonzero $\gamma $ to retain the coupling to SCR.

\begin{figure}
\centering
\includegraphics{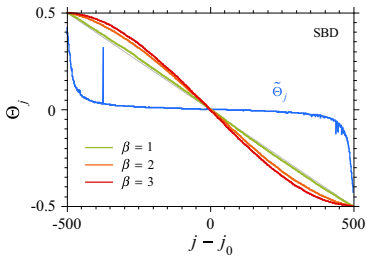}
\caption{Probe temperature ${\Theta _j}$ in the weak-coupling limit for $\beta = 1$, 2 and 3 in comparison with kinetic temperature ${\tilde \Theta _j}$ in the absence of SCR for the system with SBD of size $N = 1000$. The reference point ${j_0} = (N + 1)/2$ is used to symmetrize these plots. The thin gray line shows the temperature profile obeying Fourier's law as a reference.}
\label{fig:6}
\end{figure}

Figure \ref{fig:6} shows the numerical results obtained for SBD in the weak-coupling limit.
As seen in the figure, the probe temperature ${\Theta _j}$ does not coincide with the kinetic temperature ${\tilde \Theta _j}$.
The discrepancy relates to the presence or absence of local equilibration processes.
It is also noticed that ${\Theta _j}$ appreciably depends on the reservoir spectrum specified by $\beta $.
These phenomena are thought to be an observer effect in that an act of observation necessarily changes the physical situations of the object, and the result of observation depends on the manner of observation.
For an Ohmic reservoir with $\beta = 1$, ${\Theta _j}$ shows an almost linear profile, signaling that Fourier's law is recovered in this case.
However, this does not contradict the prevailing conjecture, since total momentum conservation is violated for $\beta = 1$, as implied from previous studies \cite{ref:29, ref:30}.
It is worth noting that the recovery of Fourier's law is nontrivial since heat current under a given bias is unaffected by SCR in the weak-coupling limit.
For the other types of disorders, the temperature profiles are nonlinear regardless of $\beta $, indicating the breakdown of Fourier's law (not shown).
To summarize, local temperatures are anomalous in a disordered chain in the absence of SCR, whereas a normal temperature profile is recovered for SBD by contact with an Ohmic SCR even in the weak-coupling limit.
This confirms that mechanisms leading to local thermal equilibrium and breaking total momentum conservation are essential for normal transport.

To corroborate the arguments in this paper, it may be instructive to consider a theoretical expression for an Ohmic reservoir with $\beta = 1$.
In this case, it is easy to find that ${m_j}{G_{j\nu }} = m\gamma {\mathcal{N}_{j\nu }}$ and thereby the kinetic and probe temperatures are simply interrelated by ${\tilde \theta _j} = {\mathcal{N}_j}{\theta _j}$ in the $\theta  \to \infty $ limit. Combining this relation and Eq. (\ref{eq:10}), the asymptotic profile of probe temperatures is explicitly expressed as
\begin{equation*}
{\theta _j} = {\theta _L} - \frac{{j - 1}}{{N - 1}}({\theta _L} - {\theta _R}) ,
\end{equation*}
for $\alpha  = 1$, giving a quantitative explanation of the linear profile shown in Fig. \ref{fig:6}.

Finally, we explain heat transport in disordered chains expected at finite temperatures.
The integrands in Eqs. (\ref{eq:1}) and (\ref{eq:3}) contain a temperature-dependent factor $\eta \equiv \tfrac{{\partial f}}{{\partial \theta }}\omega $, which satisfies $\lim_{\omega \to 0} \eta = 1$ and monotonically decreases with $\omega $ at finite temperatures.
The relevant thermal cutoff is evaluated to be $3\theta $.
Therefore, the conclusions derived in the $\theta \to \infty $ limit are reasonably justifiable even at finite temperatures as long as ${\omega _N} \ll 3\theta $.
Note that this criterion is necessarily fulfilled in the asymptotic limit $N \to \infty $.

\section{Summary}
\label{sec:4}

In this paper, we have revisited thermal transport in disordered harmonic chains in detail.
For SBD, the finite-size thermal conductivity scales normally with the system size.
However, kinetic temperatures described by the local number of states coupling to external heat reservoirs are anomalous since they exhibit a nonlinear profile in the interior of the system.
Both results are consistently explained in terms of the localization length at low frequencies.
The associated argument derives a generic formula describing the asymptotic profile of local temperatures in a disordered harmonic chain.
A linear temperature profile following Fourier's law is recovered by contact with an Ohmic SCR even in the weak-coupling limit.
This verifies that mechanisms leading to local thermal equilibrium and breaking total momentum conservation are necessary for Fourier transport in low dimensions.

\appendix
\section{Correlation Functions}
\label{appendix:A}

In this Appendix, we derive Eq. (\ref{eq:2}) by introducing two NEGFs \cite{ref:20, ref:21, ref:22} defined as
\begin{eqnarray*}
g_{jj'}^ > (t,t') & = & - i\left\langle {{q_j}(t){q_{j'}}(t')} \right\rangle ,\\
g_{jj'}^ + (t,t') & = & - i\left\langle {[{q_j}(t),{q_{j'}}(t')]} \right\rangle \theta (t - t') ,
 \end{eqnarray*}
where $t$ denotes the time variable, and $\theta (t)$ is the Heaviside step function.
In stationary situations, these two-time correlation functions depend only on the time difference $\tau  = t - t'$, and are Fourier transformed into $g_{jj'}^{ > , + }(\omega ) = \int_{ - \infty }^\infty  {d\tau {e^{i\omega \tau }}g_{jj'}^{ > , + }(\tau )} $.
The retarded Green's function $g_{jj'}^ + (\omega )$  in Fourier space is expressed as
\begin{equation*}
{g^ + }(\omega ) = \frac{1}{{M{\omega ^2} - K - {\Sigma ^ + }(\omega )}} ,
\end{equation*}
in matrix notation, where ${M_{jj'}} = {m_j}{\delta _{jj'}}$ is the mass matrix, ${K_{jj'}}$ is the force constant matrix, and ${\Sigma ^ + }(\omega ) = \sum\nolimits_\nu  {\Sigma _\nu ^ + (\omega )} $ is the retarded self-energy matrix.
In the two-terminal geometry, the self-energy stems from lead $\nu  \in \{ L,R\} $ so that the matrix element ${(\Sigma _\nu ^ + )_{jj'}}$ is nonzero only for $j = j' = \nu $.
The greater Green's function $g_{jj'}^ > (\omega )$ obeys the Keldysh equation
\begin{equation*}
{g^ > }(\omega ) = {g^ + }(\omega ){\Sigma ^ > }(\omega ){g^ - }(\omega ) ,
\end{equation*}
where ${g^ - }(\omega ) = {[{g^ + }(\omega )]^*}$ is the advanced Green's function, and ${\Sigma ^ > }(\omega ) = \sum\nolimits_\nu  {\Sigma _\nu ^ > (\omega )} $ is the greater self-energy.
The self-energy $\Sigma _\nu ^ > (\omega )$ is explicitly given by
\begin{equation*}
\Sigma _\nu ^ > (\omega ) =  - i [{f_\nu }(\omega ) + 1]{\Gamma _\nu }(\omega ) ,
\end{equation*}
and relates to the linewidth function ${\Gamma _\nu }(\omega ) =  - 2\operatorname{Im}\Sigma _\nu ^ + (\omega )$.
Following the NEGF formalism summarized above, we formulate the equal-time velocity-velocity correlation as
\begin{equation*}
\begin{split}
\left\langle {{{\dot q}_j}(t){{\dot q}_j}(t)} \right\rangle  &= i\mathop {\lim }\limits_{t' \to t} \frac{{{\partial ^2}}}{{\partial t\partial t'}}g_{jj}^ > (t,t') \\ &= \frac{i}{{2\pi }}\int_{ - \infty }^\infty  {d\omega {\omega ^2}g_{jj}^ > (\omega )}  \\ &= \frac{1}{{2\pi }}\sum\limits_\nu  {\int_{ - \infty }^\infty  {d\omega ({f_\nu } + 1){\omega ^2}{\Gamma _\nu }{{\left| {g_{j\nu }^ + } \right|}^2}} }  \\ &= \frac{1}{\pi }\sum\limits_\nu  {\int_0^\infty  {d\omega ({f_\nu } + \frac{1}{2}){\omega ^2}{\Gamma _\nu }{{\left| {g_{j\nu }^ + } \right|}^2}} } . 
\end{split}
\end{equation*}
Here, we use the symmetry relations; ${f_\nu }(\omega ) =  - {f_\nu }( - \omega ) - 1$, ${\Gamma _\nu }(\omega ) =  - {\Gamma _\nu }( - \omega )$, and ${g^ \pm }(\omega ) = {[{g^ \pm }(\omega )]^t} = {g^ \mp }( - \omega )$, where ${A^t}$ denotes the transpose of $A$.
In the last two lines, ${\Gamma _\nu }$ represents the diagonal element ${({\Gamma _\nu })_{\nu \nu }}$.
It is easy to see that the derived formula accounts for Eq. (\ref{eq:2}), in which ${g^ + }(\omega )$ is written simply as $g(\omega )$ by omitting the superscript +.

\section{Weak-Coupling Limit}
\label{appendix:B}

\begin{figure}
\centering
\includegraphics{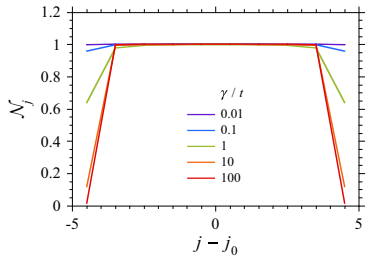}
\caption{Local number of states ${\mathcal{N}_j}$ for an ordered chain of size $N = 10$ coupling to Ohmic reservoirs at both ends. The coupling strength is varied as $\gamma /t = 0.01$, 0.1, 1, 10 and 100. The reference point ${j_0} = (N + 1)/2$ is used to symmetrize these plots. In the calculation, the reservoir bandwidth is taken as $2t$.}
\label{fig:7}
\end{figure}

To demonstrate how the system-reservoir coupling affects the local number of states ${\mathcal{N}_j}$ in the two-terminal geometry, we consider an ordered harmonic chain in contact with Ohmic reservoirs at both ends.
Figure \ref{fig:7} summarizes ${\mathcal{N}_j}$ calculated for various coupling strengths $\gamma$.
As clearly seen in the figure, ${\mathcal{N}_j}$ tends to vanish in the vicinity of the boundaries as $\gamma  \to \infty $.
On the other hand, ${\mathcal{N}_j}$ is uniform and fixed at unity for a small enough $\gamma$.
In terms of Eq. (\ref{eq:6}), this observation implies that $\tilde \theta _j^{(0)} = \theta $ in the weak-coupling limit, and does not contradict the previous results \cite{ref:2, ref:4, ref:9, ref:32}.
It should be noted that in this limit, no heat current flows across the system regardless of a temperature bias.

\bibliography{ref}
 
\end{document}